\documentclass[a4paper]{article}

\usepackage{INTERSPEECH2020}

\title{Streaming ResLSTM with Causal Mean Aggregation for Device-Directed Utterance Detection}
\name{Xiaosu Tong, Che-Wei Huang, Sri Harish Mallidi, Shaun Joseph, Sonal Pareek, Chander Chandak, Ariya Rastrow, Roland Maas}
\address{
  Amazon, USA
}
\email{\{tongx, cheweh, mallidih, josshaun, spareek, cchach, arastrow, rmaas\}@amazon.com}

\begin{document}

\maketitle
\begin{abstract}
In this paper, we propose a streaming model to distinguish voice queries intended for a smart-home 
device from background speech. The proposed model consists of multiple CNN layers with residual 
connections, followed by a stacked LSTM architecture.
The streaming capability is achieved by using unidirectional LSTM layers and a causal mean aggregation 
layer to form the final utterance-level prediction up to the current frame.
In order to avoid redundant computation during online streaming inference, we use a caching 
mechanism for every convolution operation. 
Experimental results on a device-directed vs. non device-directed task show that the proposed model 
yields an equal error rate reduction of 41\% 
compared to our previous 
best model on this task. Furthermore, we show that the proposed model is able to accurately 
predict earlier in time compared to the attention-based models.

\end{abstract}
\noindent\textbf{Index Terms}: speech recognition, human-computer interaction, computational paralinguistics

\section{Introduction}

The smart-home devices such as Amazon Echo, Google Home, etc. are often used in challenging acoustic 
conditions, such as a living room with multiple 
talkers and background media speech. In these situations, it is crucial for the device to respond only 
to the intended (referred to as device-directed (DD)) and ignore unintended (referred to as non 
device-directed (ND))
speech. We refer to ``device-directed speech detection'' as the binary utterance-level classification 
task, which can be tackled by a binary classifier trained with different types of features. 
Historically, two main types of features, acoustic features and features from Automatic Speech 
Recognition (ASR) decoding, are used in the studies of device-directed speech detection 
\cite{reich2011real, shriberg2012learning, yamagata2009system, lee2013using, wang2013understanding}.
First of all, acoustic features such as energy, pitch, speaking rate, duration and the corresponding
statistical summaries are considered in \cite{shriberg2012learning}. Other 
acoustic features such as multi-scale Gabor wavelets are studied in \cite{yamagata2009system}.
Secondly, features coming from ASR decoder such as ASR confidence scores and N-grams are 
also proved to be valuable for the detection task in \cite{shriberg2012learning, yamagata2009system}.
Comparing to the acoustic features, however the ASR decoder features are computationally more expensive,
and some of them may not even be available until the end of the utterance.

Our  previous  work  \cite{mallidi2018device, haung2019study} investigated the  
device-directed speech detection task and proposed a classifier that integrates multiple 
feature sources, including the acoustic embedding from a pretrained LSTM, speech
decoding hypothesis and decoder features from an ASR model, into one single device-directed model.
In this paper, we focus in particular on the task of learning utterance-level acoustic embeddings to 
improve the device-directed speech detection accuracy. We consider two aspects: a) the model topology 
and b) the aggregation method to convert a frame-wise into an utterance-level embedding.

As for aggregation methods, Norouzian et al. \cite{norouzian2019exploring} showed the attention 
mechanism applied to the frame-wise output of the 
network can improve the equal error rate (EER) performance of the classifier.
They used acoustic embedding features only and proposed a model 
topology consisting of a CNN and a bidirectional LSTM for the device-directed speech detection task.
Kao et al. \cite{kao2020comparison} compared different aggregation methods on top of the LSTM models for rare 
acoustic event classification, which is also an utterance classification task. 
The aggregation methods are applied to either the last hidden unit output 
$h_t$ or the soft label prediction $y_t$.

Besides the aggregation mechanism, different model topologies for audio classification tasks are 
studied in \cite{ford2019deep, bae2016acoustic, lim2017rare, cakir2017convolutional, guo2017attention,
hershey2017cnn}. Cak{\i}r et al. \cite{cakir2017convolutional} proposed a CRNN model structure for the
sound event detection task, which is similar to the CLDNN model topology proposed in 
\cite{sainath2015convolutional}. Since the results were evaluated at frame-level, no aggregation method 
was considered after the LSTM component. In \cite{guo2017attention}, the authors explored a CLDNN 
model with bidirectional LSTM combined with the attention aggregation for an acoustic scene 
classification task. Ford et al. \cite{ford2019deep} experimented with different ResNet \cite{he2016deep} 
structures, and concluded that a 50-layer ResNet shows the best performance 
on an audio event classification task. In \cite{bae2016acoustic}, instead of stacking the 
LSTM on top of a CNN component, the authors proposed a parallel structure of LSTM and CNN components. 
Then, the outputs of LSTM and CNN are concatenated and fed into the fully connected layers. 

In this paper, we evaluate the performance of different model topologies on the device-directedness 
task using acoustic features only and find the ResLSTM to outperform the ResNet, LSTM, or
CLDNN model structures.
Secondly, we propose a new mechanism to incorporate historical information within 
an utterance using frame-level causal mean aggregation. Compared to the attention method used in
\cite{haung2019study, norouzian2019exploring, kao2020comparison}, the causal mean aggregation 
\begin{itemize}
\item is able to generate prediction at any frame and easily be applied for online streaming with much
less computation.
\item has same performance as attention aggregation when evaluated at the end of an utterance.
\item outperforms the attention aggregation when evaluated at early time point of an utterance.
\end{itemize}

The rest of paper is organized as follows: Section 2 provides the overview of the main contribution
of this paper. The network architectures and different aggregation methods are discussed with details
in Section 3. Section 4 and 5 presents the experiments setup and correspondingly results. 
We conclude with Section 6.

\section{Model architecture}

In this section, we will discuss our network architectures and the aggregation methods.

\subsection{Model Topologies}

\begin{figure}[t]
  \centering
  \includegraphics[width=0.3\textwidth]{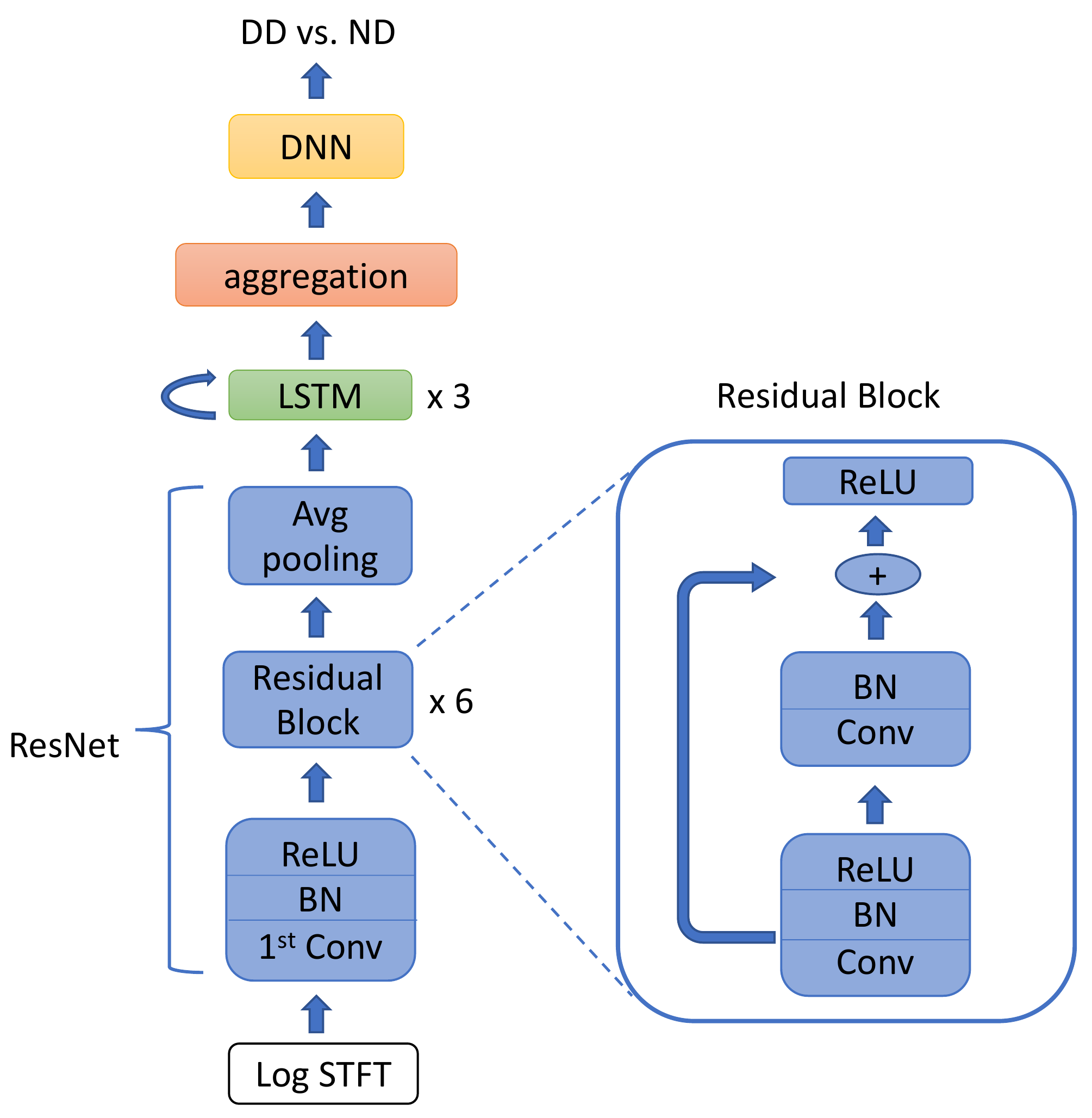}
  \caption{The structure of the ResLSTM model}
  \label{fig:model_structure}
\end{figure}

Our ResLSTM model consists of one convolutional layer and one batch norm layer followed by six 
residual blocks and one average pooling layer, as shown in Figure~\ref{fig:model_structure}. 
Each residual block has two convolution layers, two batch norm layers, and a residual connection. The 
second ReLU activation in the residual block is applied after the summation. There are 13 convolutional 
layers in total. The LSTM component has three unidirectional LSTM layers with 64 units. After the 
LSTM, there are two fully connected layers with hidden size 64.

\subsection{Aggregation}

After the frame-level embeddings $h_t$ are generated from the network, an aggregation mechanism can be 
applied to the $h_t$, and we categorize different aggregation methods into the following groups.

\subsubsection{Simple aggregation}
\label{sec:framewise}

There are two types of simple aggregation considered in this paper. First, no aggregation used at all. 
During training, we do frame-wise backpropagation on every frame with the frame-level labels which are 
obtained by repeating the utterance label. During inference, we use the embedding of the last frame
as the embedding of the entire utterance.
Second one is the global mean aggregation, which calculate the mean of the embedding of all frames as
the utterance embedding. Then, we backpropagate once for each utterance with the utterance-level
label.

\subsubsection{Attention aggregation}
\label{sec:attention}

The attention aggregation calculates the utterance-level representation as a weighted average of all
$h_t$. Similar to the global mean aggregation, it uses utterance-level label during training. 
Our previous work \cite{haung2019study} showed that the attention method has better performance 
than utterance-level embedding with global mean aggregation and frame-wise embeddings without 
any aggregation. 

\subsubsection{Causal mean aggregation}
\label{sec:causal}

The drawback of the attention methods used in the previous work 
\cite{haung2019study, norouzian2019exploring, kao2020comparison, ford2019deep} is that the attention 
weights $w_t$ 
for every frame are calculated once all frames are available, which is not feasible for online streaming 
tasks. Instead of generating the utterance-level representation at the end of the utterance, we 
generate frame-level representation by aggregating the past frames. Specifically, we average over all previous 
$h_i, i\le t$ until current time point $t$ as the representation of the $t$th frame. We call 
this the causal mean aggregation at frame-level:
\begin{equation}
  s_t = \frac{1}{t} \sum_{i=1}^{t} h_i = \frac{t-1}{t} \cdot s_{t-1} + \frac{1}{t} h_t
  \label{eq1}
\end{equation}

\begin{figure}[t]
  \centering
  \includegraphics[width=0.6\linewidth]{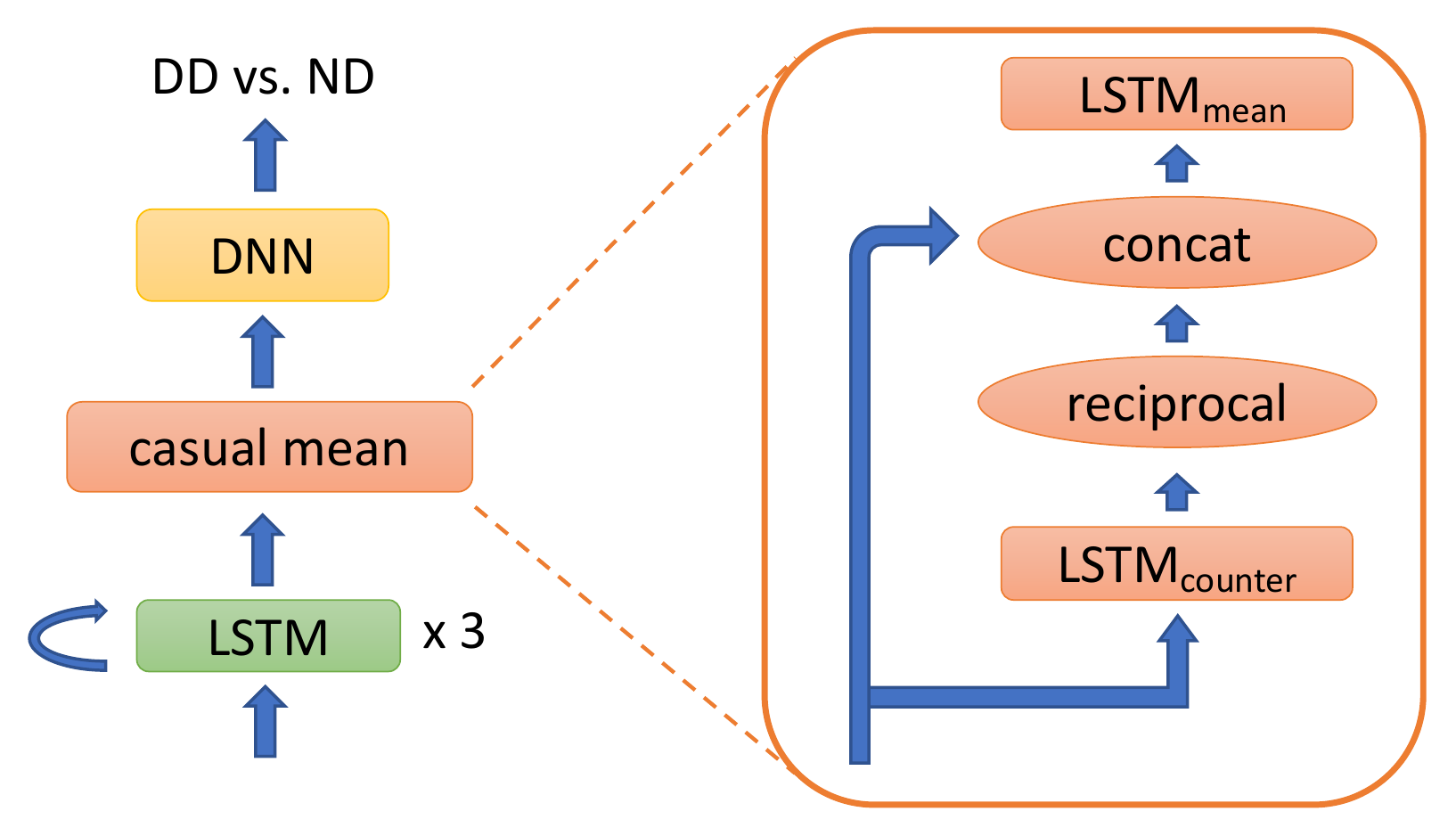}
  \caption{The implementation of causal mean aggregation for online streaming}
  \label{fig:casual_mean}
\end{figure}

During online inference, we implement the causal mean aggregation 
with a counter and a mean operation to express the logic as part of the neural network 
model definition in order to hide it from the inference engine. We find it convenient to use the 
LSTMs for this, $LSTM_{counter}$ and $LSTM_{mean}$. The LSTM structure, with state values, naturally 
allows to “side-loading” the frame count to both $h_t$ and $s_{t-1}$. 
The $LSTM_{counter}$ is for the frame counting, and the
$LSTM_{mean}$ is used for summation and division, which is shown in the Figure~\ref{fig:casual_mean}.
The two LSTM components have only one layer with fixed weights shown in Equation~\ref{eq3} and 
Equation~\ref{eq4}, respectively.
All the activation function $\sigma$ in the $LSTM_{counter}$ and $LSTM_{mean}$ components are the 
LeakyReLU with $\alpha = 1$. The LSTM gates and weights are set as follows:
\begin{equation}
\begin{split}
W_f & = 0, U_f = 0, b_f = 1 \Rightarrow f_t =1 \\
W_i & = 0, U_i = 0, b_i = 1 \Rightarrow i_t =1 \\
W_o & = 0, U_o = 0, b_o = 1 \Rightarrow o_t =1 \\
W_c & = 0, U_c = 0, b_c = 1, c_0 = 0 \\
c_t & = f_t \cdot c_{t-1} + 1 \cdot ( 0 \cdot h_t + 0 \cdot h'_t + 1) = c_{t-1} + 1 \\
h'_t & = 1 \cdot c_t = h'_{t-1} + 1 = t 
\end{split}
\label{eq3}
\end{equation}
where $W$, $U$, and $b$ are weights and bias in the forget gate ($f$), input gate ($i$), output gate 
($o$), cell input ($c$) in the $LSTM_{counter}$, respectively.
$h_t$ is the output of the original LSTM component, and $h'_t$ is the output of the $LSTM_{counter}$
component. Then, we concatenate the reciprocal of $h'_t$ with the original $h_t$ as 
$[h_t, \frac{1}{h'_t}]$ and feed into the $LSTM_{mean}$. Let's assume the dimension of 
$h_t$ is $d$, and the $LSTM_{mean}$ component has one LSTM layer with $d$ hidden units.
\begin{equation}
\begin{split}
W'_f & = [0]_{d\times (d+1)}, U'_f = [0]_{d\times d}, b'_f = [1]_{d \times 1} \Rightarrow f'_t = [1]_{d \times 1} \\
W'_i & = [0]_{d\times (d+1)}, U'_i = [0]_{d\times d}, b'_i = [1]_{d \times 1} \Rightarrow i'_t = [1]_{d \times 1} \\
W'_o & =\begin{pmatrix}
[0]_{d \times d} & [1]_{d \times 1}
\end{pmatrix},
U'_o = [0]_{d \times d},
b'_o = [0]_{d \times 1} \\
W'_c & = 
 \begin{pmatrix}
  I_{d \times d} & [0]_{d \times 1}
 \end{pmatrix},
U'_c = [0]_{d \times d}, b'_c = [1]_{d \times 1} \\
c'_t & = c'_{t-1} + h_t, o'_t = [\frac{1}{t}]_{d \times 1} \\
s_t  & = o'_t \circ c'_t = [\frac{t-1}{t}]_{d \times 1} \circ s_{t-1} + [\frac{1}{t}]_{d \times 1} \circ h_t
\end{split}
\label{eq4}
\end{equation}
where $W'$, $U'$, and $b'$ are weights and bias in the forget gate ($f$), input gate ($i$), 
output gate ($o$), cell input ($c$) in the $LSTM_{mean}$, respectively. The $\circ$ is the element-wise product,
$[k]_{d\times d}$ is a matrix with all elements equal to $k$ and dimension $d\times d$.
Similar to the idea showed in \cite{kao2020comparison}, we also move the aggregation 
component after the DNN and apply it to the $y_t$ instead of the $h_t$.

The LSTM is not the only choice to the frame counter in our implementation. A one-layer RNN
with LeakyReLU $\alpha = 1$ can be used to replace the $LSTM_{counter}$:
\begin{equation}
  h'_t = W''h_t + U''h'_{t-1} + b''
  \label{eq6}
\end{equation}
where $h$ is the output of the original LSTM component, $W'' = 0$, $U'' = 1$, $b'' = 1$, and $h'$
is initialized with 0.
Same as $LSTM_{counter}$, the output of the RNN, $h'_t$ is also the frame index $t$.
However, the RNN cannot exactly converge to mimic the $LSTM_{mean}$ because the required weight values 
($W'' = \frac{1}{t}$ and $U'' = \frac{t-1}{t}$ in Equation~\ref{eq6}) are time-dependent, which 
cannot be achieved by an RNN.

\subsubsection{RNN aggregation}
\label{sec:rnn}

In the previous section, we showed how to calculate the embedding at each frame by causal mean 
aggregation, and potentially use LSTM or RNN as the frame counter in our implementation.
Alternatively, one can use a trainable RNN layer as a different aggregation 
method besides the casual mean to get the aggregated embedding:

\begin{equation}
  s_t = \sigma (W^{T}h_t + U^{T}s_{t-1})
  \label{eq5}
\end{equation}
where $W$ and $U$ are the weights of the representation of the current frame and historical 
cumulation. The bias term $b$ is set to be 0. Instead of having the weight as fixed or predefined 
hyper parameter related to $t$ only, we use a one-layer RNN network to learn the weights for us.
But the RNN layer potentially suffers from the gradient vanishing problem over time, which we will 
see in the result session.

\subsection{Streaming CNN layer}

\begin{figure}[t]
  \centering
  \includegraphics[width=0.9\linewidth]{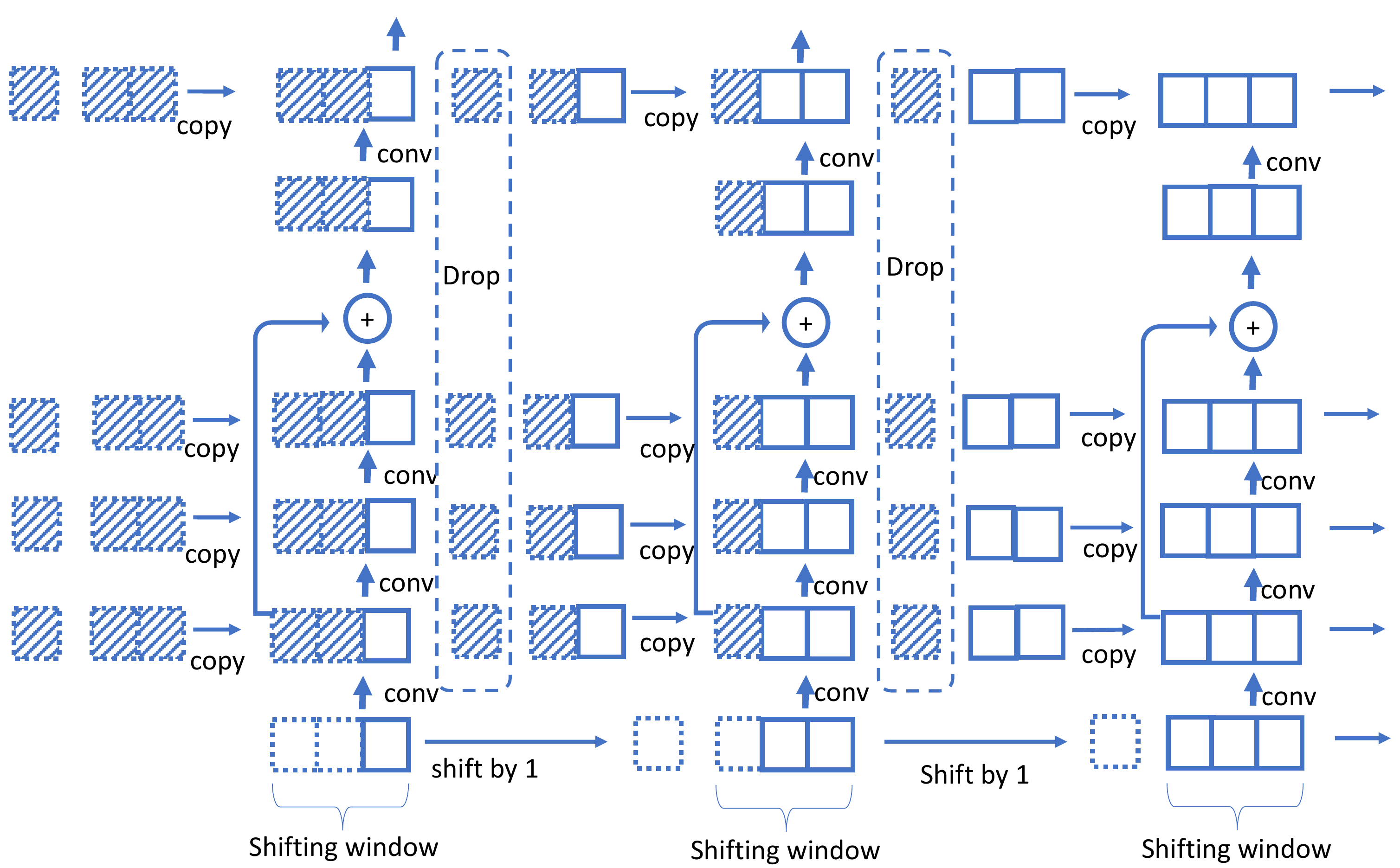}
  \caption{Streaming convolutional operation. The shaded squares represent the initialized ``previous''
  zero outputs at the first frame. The squares with a dash frame represent the frames of padded zeros, and 
  the squares with a solid frame represent the frames of real input.}
  \label{fig:model_inference}
\end{figure}

In order to enable the model with convolutional operations for online streaming, we use a sliding 
window over the input of each convolutional layer, shifting in the time dimension during inference \cite{streamcnn}. 
As the window is shifting to the right one frame at a time, we drop the oldest computation output from 
previous window, then cache 
and feed the rest of output into the next window of the same convolutional layer, which avoids wasting 
computes on redundant computations.
We initialize the ``previous'' output at the first frame to zeros. As shown in 
Figure~\ref{fig:model_inference}, we use one
convolutional layer and one residual block with the first three frame inputs as
an example for the online inference. For simplicity, we graph the the frequency dimension with size one.

\section{Experiments}

We use real recordings of natural human interactions with voice-controlled far-field devices for 
training and testing the models. The training data consists of $4,000$ hours of audio data 
comprised of 6M utterances. 4M of the utterances are device-directed examples and the rest of 2M are non 
device-directed examples. The testing data consists of 35,000 utterances.

The ResLSTM model has a kernel size as $3 \times 3$, and stride is $2 \times 1$ in which the time dimension 
stride is always 1. The output channel size of the first convolution layer is 8, and the following 6 residual 
blocks have $8, 8, 16, 16, 32, 32$ as the corresponding output channel sizes. After the last average
pooling layer, we flat out the frequency dimension with the channel dimension, and feed the outputs
to the LSTM.
We compared our ResLSTM model with LSTM, ResNet, and CLDNN models individually where we fix the aggregation
component to be attention. Two LSTM 
models are considered here. The LSTM-S has 3 LSTM layers of 64 units which is used in 
\cite{haung2019study}. The LSTM-L has 5 layers of 128 units, which is comparable to the ResLSTM 
model in terms of the number of parameters. The ResNet only model is similar to the one in the
\cite{hershey2017cnn}. We keep most of the ResNet50 \cite{hershey2017cnn} setup the same except the
following changes based on our preliminary experiments. First, we set all kernel sizes and strides to 
be $3 \times 3$ and $2 \times 1$, respectively. We also remove the max pooling layer after the first 
convolutional layer. Second, we reduce the channel sizes
to be $16, 32, 64, 128$ in each residual block. The CLDNN model we used is similar to the one used 
in \cite{guo2017attention}. It has 2 convolutional layers followed by one max pooling layer, and 
5 LSTM layers with 128 units. All the convolutional layers are followed by a batch norm layer. 

All our models are trained on the 256-dimensional log energy of short-time Fourier transform (log-STFT256)
features. For global aggregation methods, such as attention
and mean aggregation, the utterance label is used for loss calculation. We use Adam optimizer 
with the default setting \cite{torchoptimizer} to minimize the cross-entropy loss. We use low frame 
rate input which has 30ms for each frame. We truncated the input audio at 300 frames (9 seconds) length.

During the training, we feed the entire utterance input to the network. In order to match the training 
with the online streaming inference, we pad ($kernel\_size - 1$) on the left
side of time dimension of the input for every convolutional layer during training. Therefore, all 
the convolutional 
layers only see their corresponding inputs from the past but not future frames. We specify the stride 
in the time dimension of all convolutional layers to be 1.

\section{Results}

We first compare the performance across different model topologies. In the Table~\ref{tab:model_topo},
we include AUC (area under curve), EER (equal error rate), and ACC (accuracy) as our performance metrics. 
We also show the number of parameters of each model in the Table~\ref{tab:model_topo}
We use the results of LSTM-S model as the baseline. Increasing the width and depth of the LSTM-S to 
LSTM-L does reduce the EER 
by $22.6\%$
and the number of parameters is increased from 0.3M to 1M.
The CNN component in the CLDNN on top of the LSTM-L improves the EER by 
$30.0\%$. We also find simply adding
more CNN layers in the CLDNN structure does not help to improve the performance on the test dataset. 
The ResNet only model has 50 convolutional layers, and it 
improves the EER by $38.2\%$ relatively comparing to the baseline. But the number of 
parameters is about 1.5M, which is larger than other model topologies.
Finally the ResLSTM model, which has 0.9M parameters, improves the EER the most by $41.1\%$.

\begin{table}[t]
\caption{Performance of different model topology with attention aggregation}
\label{tab:model_topo}
\centering
\begin{tabular}{ccccc}
\toprule
topology  & AUC  & EER  & ACC & Para\\
\midrule
LSTM-S    & --       & --        & --      & 0.3M \\
LSTM-L    & +7.6\%  & -22.6\%  & +5.8\% & 1.0M \\
CLDNN     & +9.7\%  & -30.0\%  & +6.0\% & 1.1M \\
ResNet    & +11.9\% & -38.2\%  & +8.8\% & 1.5M \\
ResLSTM   & +12.2\% & -41.1\%  & +8.7\% & 0.9M \\
\bottomrule
\end{tabular}
\end{table}

Next, we fix the model topology to be the ResLSTM and compared different aggregation methods
including frame-level training without any aggregation \ref{sec:framewise}, utterance-level attention
\ref{sec:attention} and global mean, causal mean \ref{sec:causal} and one layer 
RNN \ref{sec:rnn}. We use the results of the ResLSTM without aggregation as 
the baseline. We applied the causal mean aggregation on either the 
LSTM output $h_t$ or the prediction 
output from the DNN $y_t$. Results are shown in the Table~\ref{tab:pooling}. As expected, 
the performance of global mean aggregation and attention method improves the EER by
$7.8\%$ and $9\%$, respectively,
which matched the finding in our previous work \cite{haung2019study}. 
The two models with frame-level causal mean aggregation show similar EER performance, which 
improves the EER by $8.5\%$ and $7.8\%$. We conclude that the model with
causal mean aggregation can achieve similar performance as model with attention when evaluate at the 
end of utterances.
Since there is no significant performance difference between the two causal mean methods, we will
use causal mean aggregation on the $h_t$ for the rest of the paper.
Using one layer RNN with $tanh$ activation function slightly improves the 
EER by $0.6\%$ comparing to the baseline. RNN with ReLU activation function performed even worse, 
increases the EER by $2.4\%$ comparing to the baseline.
We believe this is due to the gradient vanish issue of the RNN layer over time.

\begin{table}[!t]
\caption{ResLSTM model with different aggregation methods}
\label{tab:pooling}
\centering
\begin{tabular}{cccc}
\toprule
aggregation method  & AUC  & EER  & ACC \\
\midrule
frame-wise            & --  & --  & --  \\
global mean           & +1.3\% & -7.8\% & +1.0\%  \\
attention             & +1.3\% & -9.0\% & +0.6\% \\
causal mean on $h_t$  & +1.3\% & -8.5\% & +1.1\% \\
causal mean on $y_t$  & +1.5\% & -7.8\% & +1.7\% \\
RNN-ReLU              & 0\%    & +2.4\% & -1.0\% \\
RNN-tanh              & +0.2\% & -0.6\% & 0\% \\
\bottomrule
\end{tabular}
\end{table}

Instead of evaluating the prediction results at the end of utterance, we also compared the causal mean 
aggregation to the attention by evaluating the prediction at early frames in an utterance. In 
Table~\ref{tab:pooling2}, we evaluate the model performance in the first several seconds of each
utterance. The causal mean 
always performance better than the attention method in terms of EER, especially when evaluating at
the first two seconds. Moreover, in
Table~\ref{tab:pooling3}, we compare the two aggregation methods by evaluating the prediction 
results at different portions of each utterance. For example, $0.5L$ means the middle of an 
utterance. The causal mean aggregation method still consistently outperforms the attention method. 
Especially evaluating at middle of the utterance, 
it reduces the EER by $16\%$ comparing to the attention method.

\begin{table}[!t]
\caption{EER at different relative time point of the utterance. $L$ is the full length of an utterance}
\label{tab:pooling3}
\centering
\resizebox{\columnwidth}{!}{
\begin{tabular}{cccccc}
\toprule
aggregation method  & $0.5L$  & $0.6L$ & $0.7L$ & $0.8L$  & $L$ \\
\midrule
attention                & -- & -- & -- & -- & -- \\
causal mean on $h_t$     & -16.0\% & -13.3\% & -7.6\% & -3.6\% & +0.6\% \\
\bottomrule
\end{tabular}
}
\end{table}

This robustness property of the causal mean aggregation method is critical for streaming ASR 
applications for two reasons: Firstly, in practice, the end of the utterance is determined by a 
separate end-of-utterance detector (aka, end-pointer) for the purpose of ASR and can therefore vary 
significantly from utterance to utterance. Secondly, depending on the application, an early DD/ND 
decision can be desirable in order to take action prior to reaching the end of the utterance.

\begin{table}[!t]
\caption{EER at different time in seconds since beginning of the utterances.}
\label{tab:pooling2}
\centering
\resizebox{\columnwidth}{!}{
\begin{tabular}{cccccc}
\toprule
aggregation method  & $1s$  & $2s$ & $3s$ & $4s$  & $5s$ \\
\midrule
attention       & -- &  -- & --  & -- & -- \\
causal mean on $h_t$     & -4.6\% & -24.7\%  & -4.1\% & -3.3\% & -2.8\% \\
\bottomrule
\end{tabular}
}
\end{table}

We also tried a causal attention aggregation method, which mask out the future frames for the attention
calculation at each frame. But the computational cost is prohibitive, since at every frame, the attention
calculation has to be repeated 
which is much more computationally expensive than causal mean aggregation. We will consider this as a 
future work to continue seeking solution to reduce the training time.

\section{Conclusions}

In this paper, we proposed a ResLSTM model with causal mean aggregation for online streaming
classification of device-directed speech detection. Experimental results showed that the ResLSTM model
topology outperforms other topologies such as LSTM, ResNet, and CLDNN. We showed how to cache 
convolutional operations for online streaming inference with CNNs. We also proposed a causal mean 
aggregation method to obtain a more robust frame-level representation,
and showed that causal mean aggregation method can achieve the same performance as the
attention aggregation method on full utterances and significantly outperforms attention when used 
for early decision making, prior to reaching the end-of-utterance.

\bibliographystyle{IEEEtran}

\bibliography{mybib}

\begin{thebibliography}{10}
\providecommand{\url}[1]{#1}
\csname url@samestyle\endcsname
\providecommand{\newblock}{\relax}
\providecommand{\bibinfo}[2]{#2}
\providecommand{\BIBentrySTDinterwordspacing}{\spaceskip=0pt\relax}
\providecommand{\BIBentryALTinterwordstretchfactor}{4}
\providecommand{\BIBentryALTinterwordspacing}{\spaceskip=\fontdimen2\font plus
\BIBentryALTinterwordstretchfactor\fontdimen3\font minus
  \fontdimen4\font\relax}
\providecommand{\BIBforeignlanguage}[2]{{%
\expandafter\ifx\csname l@#1\endcsname\relax
\typeout{** WARNING: IEEEtran.bst: No hyphenation pattern has been}%
\typeout{** loaded for the language `#1'. Using the pattern for}%
\typeout{** the default language instead.}%
\else
\language=\csname l@#1\endcsname
\fi
#2}}
\providecommand{\BIBdecl}{\relax}
\BIBdecl

\bibitem{reich2011real}
D.~Reich, F.~Putze, D.~Heger, J.~Ijsselmuiden, R.~Stiefelhagen, and T.~Schultz,
  ``A real-time speech command detector for a smart control room,'' in
  \emph{Twelfth Annual Conference of the International Speech Communication
  Association}, 2011.

\bibitem{shriberg2012learning}
E.~Shriberg, A.~Stolcke, D.~Hakkani-T{\"u}r, and L.~Heck, ``Learning when to
  listen: Detecting system-addressed speech in human-human-computer dialog,''
  in \emph{Thirteenth Annual Conference of the International Speech
  Communication Association}, 2012.

\bibitem{yamagata2009system}
T.~Yamagata, T.~Takiguchi, and Y.~Ariki, ``System request detection in human
  conversation based on multi-resolution gabor wavelet features,'' in
  \emph{Tenth Annual Conference of the International Speech Communication
  Association}, 2009.

\bibitem{lee2013using}
H.~Lee, A.~Stolcke, and E.~Shriberg, ``Using out-of-domain data for lexical
  addressee detection in human-human-computer dialog,'' in \emph{Proceedings of
  the 2013 Conference of the North American Chapter of the Association for
  Computational Linguistics: Human Language Technologies}, 2013, pp. 221--229.

\bibitem{wang2013understanding}
D.~Wang, D.~Hakkani-T{\"u}r, and G.~Tur, ``Understanding computer-directed
  utterances in multi-user dialog systems,'' in \emph{2013 IEEE International
  Conference on Acoustics, Speech and Signal Processing}.\hskip 1em plus 0.5em
  minus 0.4em\relax IEEE, 2013, pp. 8377--8381.

\bibitem{mallidi2018device}
S.~H. Mallidi, R.~Maas, K.~Goehner, A.~Rastrow, S.~Matsoukas, and
  B.~Hoffmeister, ``Device-directed utterance detection,'' \emph{arXiv preprint
  arXiv:1808.02504}, 2018.

\bibitem{haung2019study}
C.-W. Haung, R.~Maas, S.~H. Mallidi, and B.~Hoffmeister, ``A study for
  improving device-directed speech detection toward frictionless human-machine
  interaction,'' in \emph{Proc. Interspeech}, 2019.

\bibitem{norouzian2019exploring}
A.~Norouzian, B.~Mazoure, D.~Connolly, and D.~Willett, ``Exploring attention
  mechanism for acoustic-based classification of speech utterances into
  system-directed and non-system-directed,'' in \emph{ICASSP 2019-2019 IEEE
  International Conference on Acoustics, Speech and Signal Processing
  (ICASSP)}.\hskip 1em plus 0.5em minus 0.4em\relax IEEE, 2019, pp. 7310--7314.

\bibitem{kao2020comparison}
C.-C. Kao, M.~Sun, W.~Wang, and C.~Wang, ``A comparison of pooling methods on
  lstm models for rare acoustic event classification,'' 2020.

\bibitem{ford2019deep}
L.~Ford, H.~Tang, F.~Grondin, and J.~Glass, ``A deep residual network for
  large-scale acoustic scene analysis,'' \emph{Proc. Interspeech 2019}, pp.
  2568--2572, 2019.

\bibitem{bae2016acoustic}
S.~H. Bae, I.~Choi, and N.~S. Kim, ``Acoustic scene classification using
  parallel combination of lstm and cnn,'' in \emph{Proceedings of the Detection
  and Classification of Acoustic Scenes and Events 2016 Workshop (DCASE2016)},
  2016, pp. 11--15.

\bibitem{lim2017rare}
H.~Lim, J.~Park, and Y.~Han, ``Rare sound event detection using 1d
  convolutional recurrent neural networks,'' in \emph{Proceedings of the
  Detection and Classification of Acoustic Scenes and Events 2017 Workshop},
  2017, pp. 80--84.

\bibitem{cakir2017convolutional}
E.~Cak{\i}r, G.~Parascandolo, T.~Heittola, H.~Huttunen, and T.~Virtanen,
  ``Convolutional recurrent neural networks for polyphonic sound event
  detection,'' \emph{IEEE/ACM Transactions on Audio, Speech, and Language
  Processing}, vol.~25, no.~6, pp. 1291--1303, 2017.

\bibitem{guo2017attention}
J.~Guo, N.~Xu, L.-J. Li, and A.~Alwan, ``Attention based cldnns for
  short-duration acoustic scene classification.'' in \emph{INTERSPEECH}, 2017,
  pp. 469--473.

\bibitem{hershey2017cnn}
S.~Hershey, S.~Chaudhuri, D.~P. Ellis, J.~F. Gemmeke, A.~Jansen, R.~C. Moore,
  M.~Plakal, D.~Platt, R.~A. Saurous, B.~Seybold \emph{et~al.}, ``Cnn
  architectures for large-scale audio classification,'' in \emph{2017 ieee
  international conference on acoustics, speech and signal processing
  (icassp)}.\hskip 1em plus 0.5em minus 0.4em\relax IEEE, 2017, pp. 131--135.

\bibitem{sainath2015convolutional}
T.~N. Sainath, O.~Vinyals, A.~Senior, and H.~Sak, ``Convolutional, long
  short-term memory, fully connected deep neural networks,'' in \emph{2015 IEEE
  International Conference on Acoustics, Speech and Signal Processing
  (ICASSP)}.\hskip 1em plus 0.5em minus 0.4em\relax IEEE, 2015, pp. 4580--4584.

\bibitem{he2016deep}
K.~He, X.~Zhang, S.~Ren, and J.~Sun, ``Deep residual learning for image
  recognition,'' in \emph{Proceedings of the IEEE conference on computer vision
  and pattern recognition}, 2016, pp. 770--778.

\bibitem{streamcnn}
Fifo buffer op, https://github.com/apache/incubator-tvm/pull/4039.

\bibitem{torchoptimizer}
Torch optimizer, www.pytorch.org.

\end{thebibliography}

\end{document}